\documentclass[final,5p,times,twocolumn]{elsarticle}

\usepackage{amssymb}
\usepackage{amsmath,amssymb,amscd,amsbsy,amsgen,amsopn,amstext,amsxtra}
\usepackage[mathscr]{eucal}

\newcommand{\s}{\:\!}
\newcommand{\m}{\;\!}
\newcommand{\ket}{\rangle}
\newcommand{\bra}{\langle}

\newcommand{\dket}{\rangle\!\rangle}
\newcommand{\dbra}{\langle\!\langle}
\newcommand{\dcket}{)\!)}

\newcommand{\1}{\mbox{1}\hspace{-0.25em}\mbox{l}}

\begin{document}

\begin{frontmatter}



\title{Square-integrable eigenfunctions in quantizing the Bateman oscillator model 
\begin{center}
\normalsize{-- Comment on ``A no-go result for the quantum damped harmonic oscillator"  
[Phys. Lett. A 383 (2019) 2836] --}
\end{center}
\vspace{-2.5mm}
}




\author[Deg]{Shinichi Deguchi}
\ead{deguchi@phys.cst.nihon-u.ac.jp}

\address[Deg]{Institute of Quantum Science, College of Science and Technology, 
Nihon University, Chiyoda-ku, Tokyo 101-8308, Japan}

\author[Fuj]{Yuki Fujiwara\corref{mycorrespondingauthor}}
\cortext[mycorrespondingauthor]{Corresponding author.}
\ead{yfujiwara@phys.cst.nihon-u.ac.jp}

\address[Fuj]{Department of Quantum Science and Technology, Graduate School of Science and Technology, 
Nihon University, Chiyoda-ku, Tokyo 101-8308, Japan}

\begin{abstract}
In a recent paper 
[Phys. Lett. A 383 (2019) 2836; arXiv:1906.05121 [quant-ph]], 
Bagarello, Gargano, and Roccati 
have claimed that no square-integrable vacuum exists in quantizing the Bateman oscillator model.  
In this paper, we rebut their claim by actually deriving the square-integrable vacuum eigenfunction 
using a common procedure. 
We see that no problems occur in quantizing the Bateman oscillator model.  
\end{abstract}

\begin{keyword}
Bateman oscillator model \sep 
Feshbach-Tikochinsky's quantization approach \sep 
Square-integrable eigenfunctions 
\end{keyword}

\end{frontmatter}



\section{Introduction}

Recently, Feshbach-Tikochinsky's quantization approach \cite{FesTik} to the Bateman oscillator model  
(or simply the Bateman model) \cite{Bateman} 
has been reformulated concisely without invoking the ${\mathit{SU}(1,1)}$ Lie algebra \cite{DFN}.  
In this reformulation, the Hamiltonian eigenvalues found earlier by Feshbach and Tikochinsky 
are derived simply and immediately. Also, the corresponding orthonormal eigenvectors are easily constructed by 
applying certain creation operators to the Bogoliubov vacuum state 
that is denoted in Ref. \cite{DFN} and in this paper as ${|\s0\dket}$ (see Eq. (\ref{10})).

More recently, in Ref. \cite{BGR}, Bagarello, Gargano, and Roccati 
have stated that there is no square-integrable 
eigenfunction corresponding to ${|\s0\dket}$, and therefore quantizing  
the Bateman model is impossible within the realm of Hilbert spaces. 
With this statement, Bagarello {\it et al.} claimed that the conclusions deduced in Ref. \cite{DFN} are wrong.

In this paper, we rebut the claim made by Bagarello {\it et al}., clarifying their mistake.    
We point out that the eigenfunction corresponding to ${|\s0\dket}$ has not been defined in Ref. \cite{BGR} 
precisely and correctly. 
We show that the {\em proper} inner product between bra and ket vectors 
leads to square-integrable eigenfunctions.  
In fact, the square-integrable vacuum eigenfunction 
is given as the proper inner product between a bra vector and ${|\s0\dket}$. 
We see that the quantization procedures established in Ref. \cite{DFN} have no problems. 
The purpose of this paper is not only to offer a rebuttal but also to 
reinforce Ref. \cite{DFN} with a study of  
the square-integrable eigenfunctions in quantizing the Bateman model.

This paper is organized as follows: 
In Section 2, we make a preliminary review of the necessary parts of Ref. \cite{DFN}.  
In Section 3, we introduce a statement given by Bagarello {\it et al.} in our own manner.  
In Section 4, we actually present the square-integrable eigenfunctions 
to contradict their statement. 
Section 5 is devoted to concluding remarks.

\section{Preliminaries} 

In Ref. \cite{DFN}, the operators $a_{i}$ and $a_{i}^{\dagger}$ are defined as usual by 
\begin{subequations}
\label{1}
\begin{align}
a_{i} &:=\sqrt{\frac{m\omega}{2\hbar}}\m \hat{x}_{i} +i \sqrt{\frac{1}{2\hbar m\omega}} \m \hat{p}_{i} \,,
\label{1a}
\\[3pt]
a_{i}^{\dagger} &:=\sqrt{\frac{m\omega}{2\hbar}} \m \hat{x}_{i} -i \sqrt{\frac{1}{2\hbar m\omega}} \m \hat{p}_{i} \,, 
\label{1b}
\end{align}
\end{subequations}
where $\hat{x}_{i}$ and $\hat{p}_{i}$ ${(\s i=1, 2 \s)}$ are, respectively, position and momentum operators 
that satisfy ${\hat{x}_{i}^{\dagger}=\hat{x}_{i}}$, ${\hat{p}_{i}^{\dagger}=\hat{p}_{i}}$, and 
the commutation relations 
\begin{align}
\left[\s \hat{x}_{i}\s,\s \hat{p}_{j} \right]=i\hbar \delta_{ij} \1 \quad (\s i, j=1, 2 \s) \,, 
\quad\: \mbox{all others}=0 
\label{2}
\end{align}
with the identity operator $\1$. 
(Here the over hat symbol is used to stress that $\hat{x}_{i}$ and $\hat{p}_{i}$ are operators. 
In Ref. \cite{DFN}, $\hat{x}_{i}$ and $\hat{p}_{i}$ were simply denoted as $x_{i}$ and $p_{i}$, respectively.)  
It follows from Eqs. (\ref{1}) and (\ref{2}) that 
\begin{align}
\left[\s a_{i}\s, \s a_{j}^{\dagger} \s \right]=\delta_{ij} \1 \,,  
\quad\: \mbox{all others}=0\,. 
\label{3}
\end{align}
In terms of $a_{i}$ and $a_{i}^{\dagger}$, the Hamiltonian operator governing the Bateman model 
at the quantum level is expressed as \cite{DFN}
\begin{align}
H=\hbar \omega \left( a_{1}^{\dagger} a_{1} -a_{2}^{\dagger} a_{2} \right) 
+i \frac{\hbar\gamma}{2m}  \left( a_{1} a_{2} -a_{1}^{\dagger} a_{2}^{\dagger} \right) . 
\label{4}
\end{align}
The vacuum state vector in the ${\big(a_{i}, a_{i}^{\dagger} \s\big)}$-system, 
$|^{\s} 0\ket$, is defined by 
\begin{align}
a_{i}\s |\s0\ket=0 \,. 
\label{5}
\end{align}

In order to reformulate Feshbach-Tikochinsky's quantization approach 
\cite{FesTik} without referring to the ${\mathit{SU}(1,1)}$ Lie algebra,  
the following operators have been considered in Ref. \cite{DFN}:  
\begin{align}
\bar{a}_{i}:=e^{\theta X} a_{i} \m e^{-\theta X} \s,
\quad\;\,
\bar{a}_{i}^{\ddagger}:=e^{\theta X} a_{i}^{\dagger} e^{-\theta X} \s,
\label{6}
\end{align}
where ${X:=a_{1} a_{2} +a_{1}^{\dagger} a_{2}^{\dagger}\s}$, and $\theta$ is a complex parameter. 
It is readily verified using Eqs. (\ref{6}) and (\ref{3}) that 
\begin{align}
\left[\s \bar{a}_{i}\s, \s\bar{a}_{j}^{\ddagger} \s \right]\!=\delta_{ij} \1 \,, 
\quad\; \mbox{all others}=0 \,. 
\label{7}
\end{align}
Equation (\ref{6}) can be written in the form of simple linear transformations \cite{DFN};  
in particular, when $\theta=\pm \pi/4$, Eq. (\ref{6}) becomes\footnote{~
The operators $\big(\bar{a}_{1}, \bar{a}_{2}, \bar{a}_{1}^{\ddagger}, \bar{a}_{2}^{\ddagger} \big)$ 
are identical to $(A_{1}, A_{2}, B_{1}, B_{2})$ defined in Ref. \cite{BGR} when $\theta=\pi/4$ 
and to $\big(B_{1}^{\dagger}, B_{2}^{\dagger}, A_{1}^{\dagger}, A_{2}^{\dagger} \big)$ when $\theta=-\pi/4$.  
} 
\begin{subequations}
\label{8}
\begin{alignat}{2}
\bar{a}_{1}&=\frac{1}{\sqrt{2}} \left( a_{1} \mp a_{2}^{\dagger} \right), 
&\quad \;
\bar{a}_{2}&=\frac{1}{\sqrt{2}} \left( \mp\, a_{1}^{\dagger} + a_{2} \right), 
\label{8a}
\\[3pt]
\bar{a}_{1}^{\ddagger}&=\frac{1}{\sqrt{2}} \left( a_{1}^{\dagger} \pm a_{2} \right), 
&\quad \;
\bar{a}_{2}^{\ddagger}&=\frac{1}{\sqrt{2}} \left( \pm\, a_{1} + a_{2}^{\dagger} \right).  
\label{8b}
\end{alignat}
\end{subequations}
In this case, $e^{\theta X}$ is non-unitary, 
and Eq. (\ref{8}) is recognized as a pseudo Bogoliubov transformation \cite{DFN}.

In terms of $\bar{a}_{i}$ and $\bar{a}_{i}^{\ddagger}$,  
the Hamiltonian operator $H$ can be expressed as 
\begin{align}
H=\hbar \omega \left( \bar{a}_{1}^{\ddagger} \bar{a}_{1} -\bar{a}_{2}^{\ddagger} \bar{a}_{2} \right) 
\pm i \frac{\hbar\gamma}{2m}  
\left( \bar{a}_{1}^{\ddagger} \bar{a}_{1} +\bar{a}_{2}^{\ddagger} \bar{a}_{2} +\1 \right) . 
\label{9}
\end{align}
As readily seen from Eq. (\ref{5}), the vector 
\begin{align}
|\s0\dket :=e^{\theta X} |\s0\ket 
\label{10}
\end{align}
satisfies 
\begin{align}
\bar{a}_{i}\s |\s0\dket=0 \,.        
\label{11}
\end{align}
Equation (\ref{11}) implies that ${|\s0\dket}$ is the vacuum state vector in   
the ${\big(\bar{a}_{i}, \bar{a}_{i}^{\ddagger} \s\big)}$-system.    
The vector ${|\s0\dket}$ is sometimes referred to as the Bogoliubov vacuum state.

\section{Statements in a recent paper} 

Now, let us review the statements made in Ref. \cite{BGR} in our own manner. 
For this purpose, we first consider the simultaneous eigenvalue equations 
\begin{align}
\bra x_1, x_2 |^{\s}\hat{x}_{i} =x_{i} \bra x_1, x_2 | \quad\; (i=1, 2) \m.  
\label{12}
\end{align}
The eigenvector $\bra x_1, x_2 |$ is specified by the eigenvalues ${x_{i}\s( \in \Bbb{R})}$. 
Using Eqs. (\ref{2}) and (\ref{12}), we can show that 
\begin{subequations}
\label{13}
\begin{align}
\bra x_1, x_2 |^{\s} a_{i} & 
=\left( \sqrt{\frac{m\omega}{2\hbar}}\m x_{i} 
+\sqrt{\frac{\hbar}{2m\omega}} \m \frac{\partial}{\partial x_{i}} \right) \bra x_1, x_2 |  \,,
\label{13a}
\\[3pt]
\bra x_1, x_2 |^{\s} a_{i}^{\dagger} &
=\left( \sqrt{\frac{m\omega}{2\hbar}}\m x_{i} 
-\sqrt{\frac{\hbar}{2m\omega}} \m \frac{\partial}{\partial x_{i}} \right) \bra x_1, x_2 |   
\label{13b}
\end{align}
\end{subequations}
(see, i.e., Ref. \cite{Sakurai}). 
Combining Eqs. (\ref{8a}) and (\ref{13}) yields 
\begin{subequations}
\label{14}
\begin{align}
& \bra x_1, x_2 |^{\s} \bar{a}_{1} 
\nonumber 
\\ 
& =\frac{1}{2}\left\{ \sqrt{\frac{m\omega}{\hbar}}\m (x_{1} \mp x_{2}) 
+\sqrt{\frac{\hbar}{m\omega}} \m \left( \frac{\partial}{\partial x_{1}} 
\pm \frac{\partial}{\partial x_{2}} \right) \right\} \bra x_1, x_2 |  \,,
\label{14a}
\\
& \bra x_1, x_2 |^{\s} \bar{a}_{2} 
\nonumber 
\\ 
& =\frac{1}{2}\left\{ \mp\sqrt{\frac{m\omega}{\hbar}}\m (x_{1} \mp x_{2}) 
\pm \sqrt{\frac{\hbar}{m\omega}} \m \left( \frac{\partial}{\partial x_{1}} 
\pm \frac{\partial}{\partial x_{2}} \right) \right\} \bra x_1, x_2 |  \,. 
\label{14b}
\end{align}
\end{subequations}
Multiplying both sides of Eqs. (\ref{14a}) and (\ref{14b}) by ${|\s0\dket}$ on the right 
and using Eq. (\ref{11}), we have
\begin{subequations}
\label{15}
\begin{align}
\left\{ (x_{1} \mp x_{2}) 
+\frac{\hbar}{m\omega} \m \left( \frac{\partial}{\partial x_{1}} 
\pm \frac{\partial}{\partial x_{2}} \right) \right\} \varphi_{0,0} (x_{1}, x_{2})=0 \,,
\label{15a}
\\[3pt] 
\left\{ (x_{1} \mp x_{2}) 
-\frac{\hbar}{m\omega} \m \left( \frac{\partial}{\partial x_{1}} 
\pm \frac{\partial}{\partial x_{2}} \right) \right\} \varphi_{0,0} (x_{1}, x_{2})=0 \,,
\label{15b}
\end{align}
\end{subequations}
where 
\begin{align}
\varphi_{0,0} (x_{1}, x_{2}) :=\bra x_1, x_2 |^{\s}0\dket \,.
\label{16}
\end{align}
The pair of Eqs. (\ref{15a}) and (\ref{15b}) is equivalent to the following pair of equations: 
\begin{subequations}
\label{17}
\begin{align}
(x_{1} \mp x_{2}) \m \varphi_{0,0} (x_{1}, x_{2})=0 \,, 
\label{17a}
\\
\left( \frac{\partial}{\partial x_{1}} 
\pm \frac{\partial}{\partial x_{2}} \right) \varphi_{0,0} (x_{1}, x_{2})=0 \,. 
\label{17b}
\end{align}
\end{subequations}
The solution of Eq. (\ref{17}) is found to be 
\begin{align}
\varphi_{0,0} (x_{1}, x_{2}) =c \delta(x_{1} \mp x_{2}) \,, \quad c \in \Bbb{C} \,.
\label{18}
\end{align}
Having obtained Eq. (\ref{18}), Bagarello {\it et al.} stated that the only solution 
of Eq. (\ref{17}) is a distribution \cite{BGR}. This statement itself is acceptable.  
However, we would like to 
emphasize that the $\varphi_{0,0} (x_{1}, x_{2})$ defined in Eq. (\ref{16}) is never recognized as  
the eigenfunction corresponding to  $|^{\s}0\dket $. 
It thus turns out that the main statement of Ref. \cite{BGR}   
``no square integrable vacuum exists" is totally incorrect.  
We are going to explain the details in the next section.

\section{Square-integrable eigenfunctions} 

Let $\mathcal{B}$ and $\mathcal{K}$ be the bra and ket spaces for 
the ${\big(a_{i}, a_{i}^{\dagger} \s\big)}$-system, respectively, and  
let $\bar{\mathcal{B}}$ and $\bar{\mathcal{K}}$ be the bra and ket spaces for 
the ${\big(\bar{a}_{i}, \bar{a}_{i}^{\ddagger} \s\big)}$-system, respectively. 
The bra space $\mathcal{B}$ is the dual space of $\mathcal{K}$, 
and the bra space $\bar{\mathcal{B}}$ is the dual space of $\bar{\mathcal{K}}$. 
The {\em proper} inner product is defined between an element of $\mathcal{B}$ and an element of $\mathcal{K}$,    
or between an element of $\bar{\mathcal{B}}$ and an element of $\bar{\mathcal{K}}$.   
The inner product between an element of $\mathcal{B}$ ($\bar{\mathcal{B}}$) 
and an element of $\bar{\mathcal{K}}$ ($\mathcal{K}$) is improper in common sense, 
although it may have a particular meaning.

An arbitrary element $|^{\s} \psi\ket$ of $\mathcal{K}$ is related to an element 
$|^{\s} \psi\dket$ of $\bar{\mathcal{K}}$ by 
${|^{\s} \psi\dket =e^{\theta X} |^{\s} \psi\ket}$. 
Because this relation is invertible, it follows that $\mathcal{K}$ and $\bar{\mathcal{K}}$ are isomorphic. 
Similarly, an arbitrary element $\bra\phi^{\s}|$ of $\mathcal{B}$ is related to an element 
$\dbra\phi^{\s}|$ of $\bar{\mathcal{B}}$ by ${\dbra\phi^{\s}|=\bra\phi^{\s}| e^{-\theta X}}$. 
This relation is also invertible, so that $\mathcal{B}$ and $\bar{\mathcal{B}}$ are isomorphic. 
We immediately see that 
\begin{subequations}
\label{19}
\begin{align}
\dbra\phi^{\s}|^{\s} \psi\dket &=\bra\phi^{\s}|^{\s} \psi\ket \,, 
\label{19a}
\\
|^{\s} \psi\dket \dbra\phi^{\s}| &=e^{\theta X} |^{\s} \psi\ket \bra\phi^{\s}| e^{-\theta X} .
\label{19b}
\end{align}
\end{subequations}
Equation (\ref{19a}) implies that the proper inner product is independent of the choice of the systems. 
Equation (\ref{19b}) reminds us of Eq. (\ref{6}).

Each of the operators $\hat{x}_{i}$, $ \hat{p}_{i}$, $a_{i}$, and $a_{i}^{\dagger}$ is defined on 
$\mathcal{B}$ and $\mathcal{K}$. 
Hence, it is evident that the Fock basis vectors 
\begin{align}
|^{\s} n_1, n_2 \ket :=\frac{1}{\sqrt{n_{1}! \, n_{2}!}} 
\Big(a_{1}^{\dagger} \Big)^{n_1} \Big(a_{2}^{\dagger} \Big)^{n_2} |\s0\ket 
\quad \,
(\s n_{i}=0, 1, 2, \ldots) 
\label{20}
\end{align}
are elements of $\mathcal{K}$, and the eigenvectors ${\bra x_1, x_2 |}$  ${(\s x_{i}\in \Bbb{R})}$ 
are elements of $\mathcal{B}$. 
(Clearly, ${|^{\s} n_1, n_2 \ket}$ are eigenvectors of $N_{i}:=a_{i}^{\dagger} a_{i}$ 
but not of the Hamiltonian operator $H$.) 
Using Eqs. (\ref{5}) and (\ref{13}), we have the proper inner products 
\begin{align}
& \bra x_1, x_2 |^{\s} n_1, n_2 \ket 
\notag
\\
&=\frac{1}{\sqrt{2^{(n_{1}+n_{2})} \m n_{1}! \, n_{2}!}} \left(\frac{m\omega}{\pi\hbar}\right)^{\! 1/2} 
\notag 
\\ 
& \quad \s \times  
H_{n_{1}} \! \left(\sqrt{\frac{m\omega}{\hbar}} x_{1} \right) 
H_{n_{2}} \! \left(\sqrt{\frac{m\omega}{\hbar}} x_{2} \right) 
\exp \left[\s -\frac{m\omega}{2\hbar} \left(x_1^2 +x_2^2 \right) \s\right] , 
\label{21}
\end{align}
where $H_{n}$ denotes the $n$th Hermite polynomial. 
Combining Eqs. (\ref{19a}) and (\ref{21}) gives 
\begin{align}
&\bar{\varphi}_{n_1, n_2} (x_1, x_2)
\notag
\\[3pt]
&=\frac{1}{\sqrt{2^{(n_{1}+n_{2})} \m n_{1}! \, n_{2}!}} \left(\frac{m\omega}{\pi\hbar}\right)^{\! 1/2} 
\notag 
\\ 
& \quad \s \times  
H_{n_{1}} \! \left(\sqrt{\frac{m\omega}{\hbar}} x_{1} \right) 
H_{n_{2}} \! \left(\sqrt{\frac{m\omega}{\hbar}} x_{2} \right) 
\exp \left[\s -\frac{m\omega}{2\hbar} \left(x_1^2 +x_2^2 \right) \s\right]   
\label{22}
\end{align}
with
\begin{align}
\bar{\varphi}_{n_1, n_2} (x_1, x_2)
:=\dbra x_1, x_2 |^{\s} n_1, n_2 \dket \,. 
\label{23}
\end{align}
Here, ${\dbra x_1, x_2 |:=\bra x_1, x_2 | e^{-\theta X}}$ and 
${|^{\s} n_1, n_2 \dket :=e^{\theta X} |^{\s} n_1, n_2 \ket}$. 
Since $\dbra x_1, x_2 |$ are elements of $\bar{\mathcal{B}}$, and 
$|^{\s} n_1, n_2 \dket$ are elements of $\bar{\mathcal{K}}$, 
it follows that $\dbra x_1, x_2 |^{\s} n_1, n_2 \dket$ are proper inner products. 
Accordingly, the square-integrable functions $\bar{\varphi}_{n_1, n_2} (x_1, x_2)$ 
are appreciated as the correct eigenfunctions corresponding to $|^{\s} n_1, n_2 \dket$.

Using Eq. (\ref{6}), $|^{\s} n_1, n_2 \dket$ can be expressed as 
\begin{align}
|^{\s} n_1, n_2 \dket &=\frac{1}{\sqrt{n_{1}! \, n_{2}!}} 
\Big(\bar{a}_{1}^{\ddagger} \Big)^{n_1} \Big(\bar{a}_{2}^{\ddagger} \Big)^{n_2} |\s0\dket 
\quad \,
(\s n_{i}=0, 1, 2, \ldots) .   
\label{24}
\end{align}
Also, from Eqs. (\ref{6}) and (\ref{13}), we have 
\begin{subequations}
\label{25}
\begin{align}
\dbra x_1, x_2 |^{\s} \bar{a}_{i} & 
=\left( \sqrt{\frac{m\omega}{2\hbar}}\m x_{i} 
+\sqrt{\frac{\hbar}{2m\omega}} \m \frac{\partial}{\partial x_{i}} \right) \dbra x_1, x_2 |  \,,
\label{25a}
\\[3pt]
\dbra x_1, x_2 |^{\s} \bar{a}_{i}^{\ddagger} &
=\left( \sqrt{\frac{m\omega}{2\hbar}}\m x_{i} 
-\sqrt{\frac{\hbar}{2m\omega}} \m \frac{\partial}{\partial x_{i}} \right) \dbra x_1, x_2 |  \,. 
\label{25b}
\end{align}
\end{subequations}
Equation (\ref{25a}) should be compared with Eq. (\ref{14}). 
We can directly derive Eq. (\ref{22}) by using Eqs. (\ref{11}), (\ref{24}) and (\ref{25}). 
It is easy to see that $|^{\s} n_1, n_2 \dket$ are eigenvectors of $H$ associated with 
Feshbach-Tikochinsky's Hamiltonian eigenvalues \cite{FesTik,DFN}, 
\begin{align}
\hbar \omega (n_1 -n_2) \pm i \frac{\hbar\gamma}{2m}  (n_1 +n_2 +1) \,. 
\label{26}
\end{align}

It is now obvious that the correct vacuum eigenfunction is the square-integrable function 
\begin{align}
\bar{\varphi}_{0,0} (x_{1}, x_{2})=
\left(\frac{m\omega}{\pi\hbar}\right)^{\! 1/2} 
\exp \left[\s -\frac{m\omega}{2\hbar} \left(x_1^2 +x_2^2 \right) \s\right]  
\label{27}
\end{align}
given as the proper inner product 
\begin{align}
\bar{\varphi}_{0,0} (x_1, x_2)
:=\dbra x_1, x_2 |\s0\dket  \,, 
\label{28}
\end{align}
not the distribution $\varphi_{0,0} (x_{1}, x_{2})=c \delta(x_{1} \mp x_{2})$. 
Since $\varphi_{0,0} (x_{1}, x_{2})$ is defined as the inner product between 
an element of $\mathcal{B}$ and an element of $\bar{\mathcal{K}}$, as in Eq. (\ref{16}), 
it is originally improper in the ordinary sense.

\section{Concluding remarks}

We have demonstrated that the square-integrable eigenfunction corresponding to 
the Bogoliubov vacuum state ${|\s0\dket}$   
actually exists in quantizing the Bateman model. The square-integrable eigenfunctions corresponding to  
$|^{\s} n_1, n_2 \dket$, namely ${\bar{\varphi}_{n_1, n_2} (x_{1}, x_{2})}$, have also been derived using 
the ordinary procedure mentioned above. 
Thus, we have to conclude that the conclusions deduced by Bagarello {\it et al.} are entirely wrong 
and there exist no problems in Ref. \cite{DFN}. 
Their mistake to yield the wrong conclusions is simply due to missing the correct definition of eigenfunctions.

In Ref. \cite{DFN}, the imaginary-scaling quantization approach to the Bateman model 
has been considered in addition to Feshbach-Tikochinsky's quantization approach. 
In the imaginary-scaling quantization approach, one obtains the Hamiltonian eigenvalues 
\begin{align}
\hbar \omega (n_1 +n_2 +1) \pm i \frac{\hbar\gamma}{2m} (n_1 -n_2) 
\label{29}
\end{align}
and their associated eigenvectors, which are denoted in Ref. \cite{DFN} as $|^{\s} n_1, n_2 \dcket$. 
The square-integrable eigenfunctions corresponding to $|^{\s} n_1, n_2 \dcket$ are 
derived in the same form as Eq. (\ref{22}).

\end{document}